\documentclass[%
amsmath,amssymb,
aps,
longbibliography
]{revtex4-1}

\usepackage{bm}
\usepackage{braket}
\usepackage{color}
\usepackage{dcolumn}
\usepackage{graphicx}
\usepackage[
colorlinks=true,
linkcolor=blue,
citecolor=blue,
urlcolor=blue
]{hyperref}
\usepackage{enumerate}

\newcommand{\erf}{\mathop{\mathrm{erf}} \nolimits}

\newcommand{\erfcx}{\mathop{\mathrm{erfcx}} \nolimits}

\begin{document}

\preprint{APS/123-QED}

%New command%%%%%%%%%%%%%%%%%%%%%%%%%%%%%%%%%%%%%%%%%%%%%%%%%%%%%%%%%%%%

%Title Data%%%%%%%%%%%%%%%%%%%%%%%%%%%%%%%%%%%%%%%%%%%%%%%%%%%%%%%%%%%%%

\title{Critical properties of dissipative quantum spin systems in finite dimensions}% Force line breaks with \\

%Author data%%%%%%%%%%%%%%%%%%%%%%%%%%%%%%%%%%%%%%%%%%%%%%%%%%%%%%%%%%%%
\author{Kabuki Takada}
\author{Hidetoshi Nishimori}
\affiliation{%
Department of Physics, Tokyo Institute of Technology, Oh-okayama, Meguro-ku, Tokyo 152-8551, Japan
}
\date{\today}% It is always \today, today,
             % but any date may be explicitly specified

%Abstract%%%%%%%%%%%%%%%%%%%%%%%%%%%%%%%%%%%%%%%%%%%%%%%%%%%%%%%%%%%%%
\begin{abstract}
We study the critical properties of finite-dimensional dissipative quantum spin systems with uniform ferromagnetic interactions. Starting from the transverse-field Ising model coupled to a bath of harmonic oscillators with Ohmic spectral density, we generalize its classical representation to classical spin systems with $O(n)$ symmetry and then take the large-$n$ limit to reduce the system to the spherical model. The exact solution to the resulting spherical model with long-range interactions along the imaginary-time axis shows a phase transition with static critical exponents coinciding with those of the conventional short-range spherical model in $d+2$ dimensions, where $d$ is the spatial dimensionality of the original quantum system.  This implies the dynamical exponent to be $z=2$. These conclusions are consistent with the results of Monte Carlo simulations and renormalization group calculations for dissipative transverse-field Ising and $O(n)$ models in one and two dimensions. The present approach therefore serves as a useful tool to analytically investigate the properties of quantum phase transitions of the dissipative transverse-field Ising and related models. Our method may also offer a platform to study more complex phase transitions in dissipative finite-dimensional quantum spin systems, which recently receive renewed interest under the context of quantum annealing in a noisy environment. 
\end{abstract}
%%%%%%%%%%%%%%%%%%%%%%%%%%%%%%%%%%%%%%%%%%%%%%%%%%%%%%%%%%%%%%%%%%%%%%%

\pacs{Valid PACS appear here}% PACS, the Physics and Astronomy
\maketitle

%%%%%%%%%%%%%%%%%%%%%%%%%%
\section{Introduction} \label{sec_introduction}
%%%%%%%%%%%%%%%%%%%%%%%%%%

Properties of dissipative quantum systems have been attracting attention for many years~\cite{RevModPhys.59.1,Weiss}. Recent surge of interest in quantum annealing has increased the significance of this problem because a real quantum annealing machine always operates under the effects of environmental noise~\cite{Perdomo-ortiz2015,Benedetti2015,Dickson2013,Smelyanskiy2015}. It is important in this context to understand how the critical phenomena of the transverse-field Ising model representing quantum annealing are modified by its coupling to a heat bath. There exist a large body of investigations on this problem~\cite{RevModPhys.59.1,Hertz1976,Cugliandolo2002,Cugliandolo2004,Cugliandolo2005,Vojta2005,Hoyos2008,Hoyos2012,Sinha2013,PhysRevLett.75.501,PhysRevLett.94.047201,Amin2008,Albash2015,Werner2005b,Werner2005c,Sperstad2010}, and it is generally established that the $d$-dimensional transverse-field Ising model with uniform ferromagnetic interactions, coupled to a bath of harmonic oscillators with Ohmic spectral density, has static critical exponents which are equal to those of the $(d+z)$-dimensional classical Ising model with the dynamical critical exponent $z\approx 2$. These results have been derived by several different methods including Monte Carlo simulations~\cite{PhysRevLett.94.047201,Werner2005b,Werner2005c,Sperstad2010} and renormalization group calculations~\cite{Pankov2004,Sachdev2004}. Nevertheless, complementary analytical approaches are desirable, not just to confirm these findings, but also to open a way toward investigations of more difficult cases including those with first-order phase transitions or with disorder in interactions in finite-dimensional systems, which are crucially relevant to the performance of quantum annealing under realistic settings~\cite{Perdomo-ortiz2015,Benedetti2015,Dickson2013,Smelyanskiy2015,Santoro2006,Das2008}.

The goal of the present paper is to shed new light on the problem of dissipation in quantum spin systems through the exact solution to the spherical model which is obtained as the $n\to\infty$ limit of the $O(n)$ generalization of the dissipative transverse-field Ising model. Although the spherical model generally has quantitatively different critical exponents than the Ising model, the former is known to successfully capture some of the essential features which are common to the latter, including the values of the upper and lower critical dimensions~\cite{Nishimori&Ortiz_OxfordBook,Stanley}. It is also important that the spherical model allows for the exact solution in finite dimensions even in the presence of long-range interactions~\cite{PhysRev.146.349}, which helps us lay a firm basis toward analytically reliable conclusions on more complex problems that may be out of reach by other methods. A similar idea to replace Ising spins by spherical variables was used in Refs.~\cite{Kopec1997,Cugliandolo2002} to study quantum spin glasses with infinite-range interactions.

This paper is organized as follows. The next section introduces the spherical model as the limit of infinitely many components of the $O(n)$ model, the $n=1$ case of which is the transverse-field Ising model coupled to a bath of harmonic oscillators. We derive the exact solution to the spherical model including the critical exponents and the explicit form of the correlation function. Conclusions are described in Sec.~\ref{sec_conclusion}. Technical details are delegated to Appendices.

%%%%%%%%%%%%%%%%%%%%%%%%%%
\section{Spherical model and its solution} \label{sec_model}
%%%%%%%%%%%%%%%%%%%%%%%%%%

We first introduce the $O(n)$ model with long-range interactions along the imaginary-time axis as a generalization of the classical representation of the transverse-field Ising model coupled to a bath of harmonic oscillators (to be called the dissipative transverse-field Ising model).  We then solve the model in the $n\to\infty$ limit, corresponding to the spherical model, to clarify the critical properties.

%%%%%%%%%%%%%%%%%%%%
\subsection{$O(n)$ model as a generalization of the dissipative transverse-field Ising model}
%%%%%%%%%%%%%%%%%%%%

Let us start from the Ising model in a transverse field coupled to a bath of harmonic oscillators.
The Hamiltonian is $H=H_\mathrm{S} \otimes 1_\mathrm{B} +1_\mathrm{S} \otimes H_\mathrm{B} +H_\mathrm{I}$, where
\begin{align}
H_\mathrm{S} & =-A\sum _{i=1}^N \sigma _i^x +BH_Z (\set{\sigma _i^z}), \label{eq_model.1} \\
H_\mathrm{B} & =\sum _{i=1}^N \sum _k \omega _{ik} b_{ik}^\dagger b_{ik} , \label{eq_model.2} \\
H_\mathrm{I} & =\sum _{i=1}^N \sigma _i^z \otimes\sum _k g_{ik} (b_{ik} +b_{ik}^\dagger ) . \label{eq_model.3}
\end{align}
Here, $A$ and $B$ are positive parameters, the three components of $\bm{\sigma}_i$ ($i=1,2,\dots ,N$) denote the Pauli operators, and $b_{ik}^\dagger$ and $b_{ik}$ are the $k$th bosonic operators at site $i$ with frequency $\omega _{ik}$. We keep the parameters $A$ and $B$ constant in this paper since we focus our attention on static phase transitions, although $A$ and $B$ are considered to be time dependent in the context of quantum annealing~\cite{PhysRevE.58.5355, Farhi2001,Santoro2006,Das2008}.
Notice that each site (or qubit) $i$ is independently coupled to a different set of harmonic oscillators. We assume the Ohmic-type dissipation with an infinite cutoff, i.e., the spectral density is given by $J_i (\omega ):=\sum _k g_{ik}^2 \delta (\omega -\omega _{ik})=\alpha\omega$~\cite{RevModPhys.59.1}, where $\alpha$ represents the coupling strength.
As for $H_Z$, we consider the Ising model on the $d$-dimensional hypercubic lattice
with nearest-neighbor ferromagnetic interactions with periodic boundaries under a uniform magnetic field,
\begin{equation}
H_Z (\set{\sigma _i^z})=-J_0 \sum _{\braket{ij}} \sigma _i^z \sigma _j^z -h_0 \sum _i \sigma _i^z , \label{eq_model.4}
\end{equation}
where $J_0 >0$ and $\braket{ij}$ denotes a nearest-neighbor pair.

Introducing the discrete imaginary-time variable $\tau =1,2,\dots ,M$ by the Suzuki-Trotter decomposition~\cite{Suzuki1976} and integrating out the bosonic degrees of freedom~\cite{Altland,Emery1974}, we can express the partition function at inverse temperature $\beta$ in terms of classical Ising variables $S_{i\tau}$. The result is a $(d+1)$-dimensional classical Ising model with long-range interactions along the imaginary-time axis,
\begin{equation}
Z=\lim _{M\to\infty} C\sum _{\set{S_{i\tau} =\pm 1}} \exp (-H_\mathrm{eff}^\mathrm{Ising}), \label{eq_model.5}
\end{equation}
where $C$ is an unimportant constant, and the effective Hamiltonian is given by
\begin{equation}
H_\mathrm{eff}^\mathrm{Ising} =-K\sum_{\langle ij\rangle,\tau}S_{i,\tau}S_{j,\tau}-K_{\perp}\sum_{i,\tau}S_{i,\tau}S_{i,\tau+1} -\frac{\alpha}{2}\left(\frac{\pi}{M}\right)^2\sum_{i,\tau\ne\sigma}\left(\sin\frac{\pi |\tau-\sigma|}{M}\right)^{-2}S_{i,\tau}S_{i,\sigma} -h\sum _{i,\tau} S_{i,\tau} \quad (S_{i,\tau}=\pm 1), \label{eq_model.6}
\end{equation}
where
\begin{equation}
K=\frac{\beta}{M} BJ_0 ,\quad K_\perp =-\frac{1}{2} \ln\tanh\frac{\beta A}{M} ,\quad h=\frac{\beta}{M} Bh_0 . \label{eq_model.9}
\end{equation}
The symbols $i$ and $j$ represent spatial coordinates running from $1$ to $N$, and $\sigma$ and $\tau$, both running from 1 to $M$, are for coordinates along the imaginary-time axis.

Following Refs.~\cite{Werner2005b,Werner2005c,Pankov2004}, we generalize the Ising model~\eqref{eq_model.6} to the $O(n)$ model, whose spin has $n$ components at each site with the normalization condition,
\begin{gather}
H_\mathrm{eff}^{(n)} =-K\sum_{\langle ij\rangle,\tau}\sum_{a=1}^n S_{i,\tau}^a S_{j,\tau}^a-K_{\perp}\sum_{i,\tau}\sum_{a=1}^n S_{i,\tau}^a S_{i,\tau+1}^a -\frac{\alpha}{2}\left(\frac{\pi}{M}\right)^2\sum_{i,\tau\ne\sigma}\left(\sin\frac{\pi |\tau-\sigma|}{M}\right)^{-2}\sum_{a=1}^n S_{i,\tau}^a S_{i,\sigma}^a -h\sum _{i,\tau} \sum_{a=1}^n S_{i,\tau}^a \notag \\
\left(\sum_{a=1}^n(S_{i,\tau}^a)^2=n,~\forall i, \tau \right), \label{eq_n-model}
\end{gather}
which would physically represent a single-electron box, resistively shunted Josephson junctions, or superconductor-to-metal transitions in nanowires~\cite{Werner2005b,Werner2005c,Werner2005d}. Another important reason for generalization to the $O(n)$ model is that the limit $n\to\infty$ is identical to the spherical model~\cite{PhysRev.86.821,Stanley1968},
\begin{gather}
H_\mathrm{eff}^\mathrm{sp} =-K\sum_{\langle ij\rangle,\tau}S_{i,\tau}S_{j,\tau}-K_{\perp}\sum_{i,\tau}S_{i,\tau}S_{i,\tau+1} -\frac{\alpha}{2}\left(\frac{\pi}{M}\right)^2\sum_{i,\tau\ne\sigma}\left(\sin\frac{\pi |\tau-\sigma|}{M}\right)^{-2}S_{i,\tau}S_{i,\sigma} -h\sum _{i,\tau} S_{i,\tau} \notag \\
\left(-\infty<S_{i,\tau}<\infty ,\quad \sum_{i=1}^N \sum_{\tau=1}^M (S_{i,\tau})^2 =NM \right). \label{spherical_model}
\end{gather}
The spherical model is exactly solvable for arbitrary dimensionality even in the presence of long-range interactions~\cite{PhysRev.146.349}.  Since the spherical model is known to have critical properties, some of which are shared by the Ising model~\cite{Stanley1968,Nishimori&Ortiz_OxfordBook,Stanley}, we study Eq.~\eqref{spherical_model} to extract useful information on the original dissipative Ising model in finite dimensions, Eq.~\eqref{eq_model.6}. Similar approaches were successfully used to understand the properties of quantum spin glasses with infinite-range interactions~\cite{Kopec1997,Cugliandolo2002}. We assume throughout this paper that $M$ and $\beta$ are large but finite since the critical properties are generally considered to be independent of these parameters, in particular in the present case of Ohmic bath spectrum~\cite{Vojta2005}.

It is straightforward to evaluate the partition function and the free energy of the spherical model in the limit of large system size $N$ by the standard method~\cite{PhysRev.86.821,Stanley1968,PhysRev.146.349,Nishimori&Ortiz_OxfordBook}:
\begin{align}
& Z\approx e^{-N\beta f(z)} , \label{eq_model.10} \\
& \frac{\beta}{M} f(z)=-\frac{1}{2} \ln 2\pi -z-\frac{h^2}{2} \frac{1}{2z-\tilde{K}} +\frac{1}{2} \int _{-\pi}^\pi \frac{d^d \bm{k}}{(2\pi )^d} \frac{1}{M} \sum _\kappa \ln (2z-\tilde{K} (\bm{k} ,\kappa )). \label{eq_model.11}
\end{align}
Here, $\tilde{K} (\bm{k} ,\kappa )$ is the Fourier transform of the coupling constants
\begin{align}
K(\bm{r} ,\rho ) & =K\chi [\mbox{$\bm{r}$ is a nearest-neighbor vector and $\rho =0$}] \notag \\
& +K_\perp \chi [\mbox{$\bm{r} =\bm{0}$ and $\lvert\rho\rvert =1$}] \notag \\
& +\alpha\left(\frac{\pi /M}{\sin (\pi\lvert\rho\rvert /M)} \right) ^2 \chi [\mbox{$\bm{r} =\bm{0}$ and $\rho\not= 0$}], \label{eq_model.7}
\end{align}
where $\bm{r}$ and $\rho$ denote the relative coordinates in the spatial and imaginary-time directions, respectively, and $\chi [A]=1$ if $A$ is true and zero otherwise. Additionally, we set $\tilde{K} =\tilde{K} (\bm{0} ,0)$. The variable $z$~($>\tilde{K} /2$) is the solution to the saddle point equation
\begin{equation}
1-\left(\frac{h}{2z-\tilde{K}} \right) ^2 =H(z):=\int _{-\pi}^\pi \frac{d^d \bm{k}}{(2\pi )^d} \frac{1}{M} \sum _\kappa \frac{1}{2z-\tilde{K} (\bm{k},\kappa )} . \label{eq_model.12}
\end{equation}
When $h=0$ and $H(\tilde{K} /2)<1$, the saddle point $z$ sticks to $\tilde{K} /2$,
and the system is in the ordered (ferromagnetic) phase. In other words, if there exists a set of parameters $(K,K_{\perp} ,\alpha )$ such that $H(\tilde{K} /2)=1$, a second-order phase transition exists. Equation~\eqref{eq_model.11} supplemented by Eq.~\eqref{eq_model.12} is the formal exact solution to the dissipative spherical model, Eq.~\eqref{spherical_model}.

%%%%%%%%%%%%%%
\subsection{Correlation functions}
%%%%%%%%%%%%%%

We are ready to calculate the correlation functions near criticality.
Since $K(\bm{r} ,\rho)=K(-\bm{r} ,-\rho )$, the Fourier transform is
\begin{equation}
\tilde{K} (\bm{k} ,\kappa )=\sum _{\bm{r} \in\mathbb{Z}^d} \sum _{\rho =-\frac{M-1}{2}}^{\frac{M-1}{2}} K(\bm{r} ,\rho )\cos (\bm{k} \cdot\bm{r} +\kappa\rho ), \label{eq_model.13}
\end{equation}
where we assume $M$ to be odd to simplify the analysis, which would not affect the results for sufficiently large $M$. The wave number along the imaginary-time axis $\kappa$ has the following values:
\begin{equation}
\kappa =\frac{2\pi\nu}{M} ,\quad \nu =-\frac{M-1}{2} ,\dots ,\frac{M-3}{2} ,\frac{M-1}{2} . \label{eq_model.14}
\end{equation}
As shown in Appendix~\ref{sec_freffint}, $\tilde{K} (\bm{k} ,\kappa )$ 
behaves for $\lvert\kappa\rvert\ll 1$ as follows:
\begin{equation}
\tilde{K} (\bm{k} ,\kappa )\approx 2\left[ K\sum _{a=1}^d \cos k_a +K_\perp \cos\kappa +\frac{\pi}{2} \alpha\left(\frac{\pi}{3} -\lvert\kappa\rvert\right)\right] , \label{eq_model.15}
\end{equation}
where $\bm{k} =(k_1 ,\dots ,k_d)$.

Because the terms satisfying $\lvert\bm{k} \rvert ,\lvert\kappa\rvert\ll 1$ dominate the integral on the right-hand side of Eq.~\eqref{eq_model.12} near the critical point $z=\tilde{K}/2$, we can correctly evaluate the asymptotic behavior of correlation functions near criticality by the approximation of $\tilde{K} (\bm{k} ,\kappa )$ for small $|\bm{k}|$ and $|\kappa|$ as
\begin{equation}
\tilde{K} -\tilde{K} (\bm{k} ,\kappa )\approx K\bm{k}^2 +K_\perp \kappa ^2 +\pi\alpha\lvert\kappa\rvert .
\end{equation}
Then the asymptotic form of the correlation function
\begin{equation}
G(\bm{r} ,\rho )=\int _{-\pi}^\pi \frac{d^d \bm{k}}{(2\pi )^d} \frac{1}{M} \sum _\kappa \frac{e^{i(\bm{k} \cdot\bm{r} +\kappa\rho )}}{2z-\tilde{K} (\bm{k} ,\kappa )} \approx\int _{-\pi}^\pi \frac{d^d \bm{k} \, d\kappa}{(2\pi )^{d+1}} \frac{e^{i(\bm{k} \cdot\bm{r} +\kappa\rho )}}{2z-\tilde{K} (\bm{k} ,\kappa )} \label{eq_model.18}
\end{equation}
can be described by the expression
\begin{equation}
G(\bm{r} ,\rho )\approx\int _{-\pi}^\pi \frac{d^d \bm{k} \, d\kappa}{(2\pi )^{d+1}} \frac{e^{i(\bm{k} \cdot\bm{r} +\kappa\rho )}}{u+K\bm{k}^2 +K_\perp \kappa ^2 +\pi\alpha\lvert\kappa\rvert} , \label{eq_model.19}
\end{equation}
where $u=2z-\tilde{K}$. Let us assume that $u>0$, i.e., the system is in the paramagnetic phase.
If $r=\lvert\bm{r} \rvert\gg\sqrt{K/u} =\max _\kappa \sqrt{K/(u+K_\perp \kappa ^2 +\pi\alpha\lvert\kappa\rvert)}$, we can evaluate the integral with respect to $\bm{k}$ by the Ornstein-Zernike formula as
\begin{equation}
G(\bm{r} ,\rho )\sim\frac{1}{K} \int\frac{d\kappa}{2\pi} e^{i\kappa\rho} \exp\left( -\sqrt{\frac{u+K_\perp \kappa ^2 +\pi\alpha\lvert\kappa\rvert}{K}} \, r\right) . \label{eq_model.20}
\end{equation}
The fact that $r$ is sufficiently large and the absolute value of the integrand is maximum at $\kappa =0$ yields
\begin{equation}
G(\bm{r} ,\rho )\sim\frac{1}{K} e^{-r/\xi} \quad (r\gg\xi ) , \label{eq_model.21}
\end{equation}
where the correlation length is
\begin{equation}
\xi =\sqrt{\frac{K}{u}} . \label{eq_model.22}
\end{equation}
Equation~\eqref{eq_model.21} with Eq.~\eqref{eq_model.22} is the exact asymptotic form of the spatial correlation function in the paramagnetic phase $u>0$.

On the other hand, the correlation function along the imaginary-time axis obeys a power law:
\begin{equation}
G(\bm{0} ,\rho )\approx\frac{\alpha}{(u\rho )^2} \quad\left(\rho\gg\frac{\pi (2K_\perp +\alpha )}{u} \right) , \label{eq_model.23}
\end{equation}
which we show in Appendix~\ref{sec_imtcorr} since the computations are somewhat involved. Because of the power decay of this correlation function, we cannot define the correlation time in the paramagnetic phase, a situation similar to the case of the dissipative transverse-field Ising model~\cite{PhysRevLett.94.047201} and the Josephson length~\cite{PhysRevLett.75.501}. It may in principle be possible to calculate the connected correlation function along the imaginary-time axis in the ferromagnetic phase, but it is in practice difficult. The connected correlation function in the ferromagnetic phase is expected to decay exponentially, from which the correlation time $\xi _\tau$ is defined and consequently the dynamical exponent $z$ is extracted from $\xi _\tau \sim\xi ^z$.  We leave it an open problem to evaluate $z$ directly along this line of analysis.

%%%%%%%%%%%%%%%%%%%%%%%%%%
\subsection{Critical Behavior} \label{sec_critbhv}
%%%%%%%%%%%%%%%%%%%%%%%%%%

We next calculate the critical exponents of the spherical model.
First, the behavior of $H(z)$ is investigated in order to extract critical properties using the large-wavelength expression of the integrand of the function $H(z)$, which is relevant to the critical behavior,
\begin{align}
H(z) & \approx\int _{-\pi}^\pi \frac{d^d \bm{k} \, d\kappa}{(2\pi )^{d+1}} \frac{1}{u+K\bm{k}^2 +K_\perp \kappa ^2 +\pi\alpha\lvert\kappa\rvert} \notag \\
& \sim\frac{1}{K_\perp} \int _0^\pi dk\, k^{d-1} \int _0^\pi d\kappa\left[\left(\kappa +\frac{\pi\alpha}{2K_\perp} \right) ^2 -\left(\frac{\pi\alpha}{2K_\perp} \right) ^2 +\frac{Kk^2 +u}{K_\perp} \right] ^{-1} . \label{eq_model.17}
\end{align}
Here the symbol $\approx$ stands for an approximation which is relevant to the asymptotic behavior, and $\sim$ denotes a similar approximation but with an unimportant constant dropped.
Since the choice of the lattice spacing along the imaginary-time axis, $\beta /M$, does not affect the universality which the Hamiltonian~\eqref{spherical_model} exhibits, we fix $\beta /M$ to a sufficiently small value to simplify the analysis. This supposition yields $1\gg \alpha/K_\perp \gg K/\alpha$. Noticing that $\left(\pi\alpha/2K_\perp \right) ^2 \gg (Kk^2 +u)/K_\perp$ because $u\approx 0$ near criticality, we obtain
\begin{equation}
H(z)\sim\int _0^\pi dk\,\frac{k^{d-1}}{2cK_\perp} \ln\frac{\left( 1+\frac{\alpha}{2K_\perp} -\frac{c}{\pi} \right)\left(\frac{1}{2} +\frac{K_\perp}{\pi\alpha} c\right)}{\left( 1+\frac{\alpha}{2K_\perp} +\frac{c}{\pi} \right)\left(\frac{1}{2} -\frac{K_\perp}{\pi\alpha} c\right)} , \label{eq_critbhv-diss.1}
\end{equation}
where
\begin{equation}
c(u,k)=\sqrt{\left(\frac{\pi\alpha}{2K_\perp} \right) ^2 -\frac{Kk^2 +u}{K_\perp}} \approx\frac{\pi\alpha}{2K_\perp} -\frac{Kk^2 +u}{\pi\alpha} . \label{eq_critbhv-diss.2}
\end{equation}
Use of $(2cK_\perp )^{-1} \sim\alpha ^{-1}$ and
\begin{equation}
\ln\frac{\left( 1+\frac{\alpha}{2K_\perp} -\frac{c}{\pi} \right)\left(\frac{1}{2} +\frac{K_\perp}{\pi\alpha} c\right)}{\left( 1+\frac{\alpha}{2K_\perp} +\frac{c}{\pi} \right)\left(\frac{1}{2} -\frac{K_\perp}{\pi\alpha} c\right)} \approx\ln\frac{\left( 1+\frac{Kk^2 +u}{\pi ^2 \alpha} \right)\left( 1-\frac{K_\perp (Kk^2 +u)}{(\pi\alpha )^2} \right)}{\left( 1+\frac{\alpha}{K_\perp} -\frac{Kk^2 +u}{\pi ^2 \alpha} \right)\frac{K_\perp (Kk^2 +u)}{(\pi\alpha )^2}} \approx -\ln\frac{K_\perp (Kk^2 +u)}{(\pi\alpha )^2} \label{eq_critbhv-diss.3}
\end{equation}
results in
\begin{equation}
H(z)\sim\frac{1}{\alpha} \int _0^\pi dk\, k^{d-1} \left( -\ln\frac{K_\perp (Kk^2 +u)}{(\pi\alpha )^2} \right) . \label{eq_critbhv-diss.4}
\end{equation}
Therefore, we conclude that
\begin{equation}
H(z=\tilde{K} /2)\sim\frac{1}{\alpha} \int _0^\pi dk\, k^{d-1} \ln\frac{(\pi\alpha )^2 G}{k^2} , \label{eq_critbhv-diss.5}
\end{equation}
where $G=(KK_{\perp})^{-1}$. This equation shows that  $H(\tilde{K}/2)$ converges if and only if $d>0$.

\begin{figure}
\centering
\includegraphics[scale=0.25]{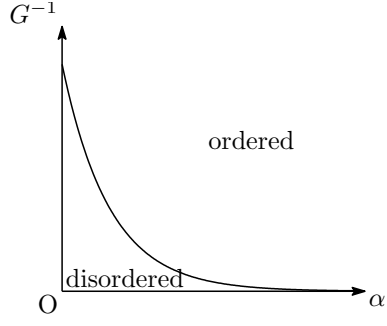}
\caption{Phase diagram of the present spherical model. For $d>0$, a phase transition occurs and the phase boundary is $G^{-1} \sim\exp (-C_d \alpha )$.} \label{fig_critbhv-diss.1}
\end{figure}

For $d>0$, we simplify Eq.~\eqref{eq_critbhv-diss.5} using the fact that $\ln G$ is dominant for small $\beta /M$:
\begin{equation}
H(\tilde{K} /2)\sim \frac{\ln G}{\alpha} . \label{eq_critbhv-diss.6}
\end{equation}
As a result, the system is in the ordered (ferromagnetic) phase if $\ln G/\alpha \lesssim C_d \iff G^{-1} \gtrsim\exp (-C_d \alpha )$ and is in the disordered (paramagnetic) phase if $G^{-1} \lesssim\exp (-C_d \alpha )$, where $C_d$ is a constant which depends only on $d$. We show the resulting phase diagram in Fig.~\ref{fig_critbhv-diss.1}.
The phase boundary $G^{-1} \sim\exp (-C_d \alpha )$ resembles those in Refs.~\cite{PhysRevLett.94.047201,Werner2005b,Sperstad2010}  for the dissipative Ising and $XY$ models. A difference is that there is no transition on the line $G^{-1}=0$ in the present case whereas the dissipative Ising model has one.  This would originate in the continuous nature of the spherical spins~\cite{Werner2005b}.

To clarify the critical behavior,
we investigate the $u$-dependence of $H(z=\tilde{K}/2)-H(z)$ for $d>0$. Changing variables to $x=Kk^2/u$ in Eq.~\eqref{eq_critbhv-diss.4} results in
\begin{equation}
H(\tilde{K} /2)-H(z)\sim\frac{1}{\alpha} \left(\frac{u}{K} \right) ^{d/2} \int _0^{K\pi ^2 /u} dx\, x^{(d-2)/2} \ln\left( 1+\frac{1}{x} \right) . \label{eq_critbhv-diss.7}
\end{equation}
Since the behavior of the integrand is
\begin{equation}
x^{(d-2)/2} \ln\left( 1+\frac{1}{x} \right)\approx\left\{
\begin{alignedat}{2}
& {-x^{(d-2)/2} \ln x} ,\quad && x\ll 1, \\
& x^{(d-4)/2} ,\quad && x\gg 1,
\end{alignedat}
\right. \label{eq_critbhv-diss.8}
\end{equation}
it is only for $d\geq 2$ that the region $x\gg 1$ dominates the integral. Consequently, we have
\begin{equation}
H(\tilde{K} /2)-H(z)\sim\frac{1}{\alpha} \times\left\{
\begin{alignedat}{2}
& \left(\frac{u}{K} \right) ^{d/2} ,\quad && 0<d<2, \\
& {-\frac{u}{K} \ln\frac{u}{K}} ,\quad && d=2, \\
& \frac{u}{K} ,\quad && 2<d.
\end{alignedat}
\right. \label{eq_critbhv-diss.9}
\end{equation}
This equation coincides with the corresponding equation of the standard classical ($d+2$)-dimensional spherical model with nearest-neighbor interactions~\cite{Nishimori&Ortiz_OxfordBook}. That is, the $d$-dimensional dissipative quantum system corresponds to the ($d+2$)-dimensional classical system.

For $0<d<2$, we solve the saddle point equation~\eqref{eq_model.12} and derive the critical exponents. If $d > 2$, one should formally insert $d=2$. In other words, the upper critical dimension is two. We fix the coupling strength $\alpha$ and concentrate on the phase transition which occurs as $G$ changes. This arrangement reflects the situation of quantum annealing. For $h=0$, Eq.~\eqref{eq_model.12} leads to $H(z)=1=H(\tilde{K}_\mathrm{c} /2)$, where $\tilde{K}_\mathrm{c}$ denotes the value of $\tilde{K}$ at the critical point $G=G_\mathrm{c}$. Using Eqs.~\eqref{eq_critbhv-diss.6} and \eqref{eq_critbhv-diss.9}, we obtain
\begin{equation}
\frac{1}{\alpha} \left(\frac{u}{K} \right) ^{d/2} \sim\frac{\ln G-\ln G_\mathrm{c}}{\alpha} \approx\frac{1}{\alpha} \frac{G-G_\mathrm{c}}{G_\mathrm{c}} \label{eq_critbhv-diss.10}
\end{equation}
and hence
\begin{equation}
2z-\tilde{K} =u\sim Kg^{2/d} , \label{eq_critbhv-diss.11}
\end{equation}
where $g=(G-G_\mathrm{c})/G_\mathrm{c}$. This solution allows us to rewrite physical quantities like the free energy expressed in terms of $u$ by the physical variable $g$. It is then straightforward to calculate the following critical exponents in a standard manner~\cite{Nishimori&Ortiz_OxfordBook}:
\begin{equation}
\alpha =\frac{d-2}{d} ,\quad \beta =\frac{1}{2} ,\quad \gamma =\frac{2}{d} ,\quad \delta =\frac{d+4}{d} . \label{eq_critbhv-diss.12}
\end{equation}
Remember that $\alpha$ and $\beta$ here denote the critical exponents of the specific heat and the magnetization, respectively, which should not be confused with the spin-bath coupling constant and the inverse temperature. Next, Eqs.~\eqref{eq_model.22} and \eqref{eq_critbhv-diss.11} yield the correlation length $\xi\sim g^{-1/d}$ and the critical exponent
\begin{equation}
\nu =\frac{1}{d} . \label{eq_critbhv-diss.13}
\end{equation}
We find that the exponents~\eqref{eq_critbhv-diss.12} and \eqref{eq_critbhv-diss.13} coincide with those for the standard classical ($d+2$)-dimensional spherical model. This result suggests that the dynamical exponent is $z=2$. In addition, we can derive the exponent $z+\eta =2$ from Eq.~\eqref{eq_model.19}, followed by $\eta =0$.

These values of critical exponents are to be compared with the results of Monte Carlo simulations of the dissipative Ising and $XY$ models in one and two dimensions~\cite{PhysRevLett.94.047201,Werner2005b,Sperstad2010} (see in particular Table I of Ref.~\cite{Werner2005b}). For example, the dissipative transverse-field Ising model in one dimension has $\nu =0.638(3)$, $\eta =0.015(20)$ and $z=1.985(15)$ according to Monte Carlo simulations~\cite{Werner2005b}, whereas our dissipative spherical model with $d=1$ has $\nu =1$, $\eta =0$ and $z=2$. The difference in $\nu$ (0.638(3) and 1) reflects the difference in this exponent between the standard classical Ising model in three dimensions ($\nu =0.6301(4)$~\cite{Pelissetto2002}) and the spherical model in three dimensions ($\nu =1$~\cite{Nishimori&Ortiz_OxfordBook,Stanley}).
It is also noticed that the $\epsilon$~($=2-d$)-expansion of critical exponents of the $O(n)$ model discussed in Refs.~\cite{Sachdev2004,Werner2005b} is in perfect agreement with our results.  For example, the $\epsilon$-expansion of the exponent $\nu$~\cite{Sachdev2004,Werner2005b} is, in the $n\to\infty$ limit,
\begin{equation}
\nu =\frac{1}{2}+\frac{1}{4}\epsilon +\frac{1}{8}\epsilon ^2+O(\epsilon^3),
\end{equation}
which is correctly reproduced by Eq.~\eqref{eq_critbhv-diss.13} with $d=2-\epsilon$.  Our approach and the renormalization group method are complementary to each other in the sense that the former is exact for any $d$ in the limit $n\to\infty$ whereas the latter is valid for any $n$ as long as $\epsilon$ is not large.

%%%%%%%%%%%%%%%%%%%%%%%%%%
\section{Conclusion} \label{sec_conclusion}
%%%%%%%%%%%%%%%%%%%%%%%%%%

We have studied the critical properties of the spherical model with long-range interactions along the imaginary-time axis, in addition to the usual nearest-neighbor ferromagnetic interactions in the spatial and imaginary-time axes. This model has been introduced as the $n\to\infty$ limit of the $O(n)$ generalization of the transverse-field Ising model coupled to a bath of harmonic oscillators with Ohmic spectral density with the ultimate goal in mind to understand the effects of noise on quantum annealing.

The solution to our dissipative spherical model has revealed that the static critical exponents in $d$ spatial dimensions completely coincide with those of the classical spherical model in $d+2$ dimensions. One of the strengths of our method lies in the exactness of the values of static critical exponents for the spherical version of the model.  If we use the standard argument that the effective dimensionality of a quantum system in $d$ spatial dimensions is $d+z$~\cite{Hertz1976,Sachdev1999}, where $z$ is the dynamical critical exponent, we conclude that the dynamical exponent is exactly equal to two. The result $z=2$ is consistent with a naive power-counting argument~\cite{Sperstad2010} using the bare propagator $k^2+|\kappa|$ of a scalar field theory corresponding to the dissipative transverse-field Ising model, in which one compares the $k^2$ term with the linear-$\kappa$ term, requiring $|\kappa|$ to be comparable to $k^z$, i.e. $|\kappa| \approx k^z$ with $z=2$. It is remarkable that this simple power-counting argument for the bare propagator of a scalar field theory turns out to be essentially exact for the present spherical model as also seen in Eq.~\eqref{eq_model.19}.

We should of course be careful to draw too strong conclusions on the original problem of the dissipative transverse-field Ising model.  Nevertheless, Monte Carlo simulations and renormalization-group calculations of the $d=1$ and $d=2$ dissipative transverse-field Ising models show that the static critical exponents are close to those of the classical Ising models in $d=3$~($=1+2$) and $d=4$~($=2+2$), respectively, and the dynamical exponent $z$ is close to two~\cite{PhysRevLett.94.047201,Werner2005b,Werner2005c,Sperstad2010,Pankov2004,Sachdev2004}. Similar results are also found in the dissipative $XY$ model in one dimension~\cite{Werner2005b,Werner2005d}. These findings are consistent with our results for the spherical model, which renders a support to the expectation that the analysis of the spherical model is a useful tool to understand the effects of dissipation on the transverse-field Ising model.

From the viewpoint of quantum annealing, the existence of a second-order phase transition does not spell a difficulty because the energy gap at such a transition point closes polynomially as a function of the system size~\cite{Sachdev1999}, which means that the computation time of quantum annealing grows only polynomially and thus the problem is easy to solve. This implies that the quantitative change of critical exponents by dissipation affects the performance of quantum annealing only quantitatively, not qualitatively, in the present system with uniform ferromagnetic interactions. More serious and practically important are the instances with a first-order phase transition and/or randomness in interactions, in particular in finite dimensions, which are the target of our work in progress.

\subsection*{Acknowledgement} %%%%%%%%%%%%%%%%%%%%%%%%%%%%%%%%%

We thank Matthias Troyer and Sergey Knysh for useful comments.
This work was funded by ImPACT Program of Council for Science, Technology and Innovation, Cabinet Office, Government of Japan, and by the JPSJ KAKENHI Grant No. 26287086.

\appendix

%%%%%%%%%%%%%
\section{Large-wavelength Behavior of the Interaction} \label{sec_freffint}
%%%%%%%%%%%%%%%%%%%%%%%%%%

Here we derive Eq.~\eqref{eq_model.15}. Although the result has been known for years, an explicit presentation of its rigorous derivation seems to be missing in references, and thus it is useful to write out the derivation for the paper to be self-contained.

The first and second terms on the right-hand side of Eq.~\eqref{eq_model.15} produced by nearest-neighbor interactions are trivial, so that we derive only the following expression:
\begin{equation}
\left(\frac{\pi}{M} \right) ^2 \sum _{\rho =1}^{\frac{M-1}{2}} \frac{\cos\kappa\rho}{\sin ^2 \frac{\pi\rho}{M}} \approx\frac{\pi}{2} \left(\frac{\pi}{3} -\lvert\kappa\rvert\right) , \label{eq_freffint.1}
\end{equation}
which is valid for $\lvert\kappa\rvert\ll 1$.
Changing the variable $\kappa$ to $\nu$ with Eq.~\eqref{eq_model.14}, the left-hand side of Eq.~\eqref{eq_freffint.1} is rewritten as
\begin{equation}
S_\nu :=\left(\frac{\pi}{M} \right) ^2 \sum _{\rho =1}^{\frac{M-1}{2}} \frac{\cos\frac{2\pi\nu\rho}{M}}{\sin ^2 \frac{\pi\rho}{M}} . \label{eq_freffint.2}
\end{equation}
Although $\nu$ can have values such that $\nu ^{-1} =o(M^0)$, we show
\begin{equation}
S_\nu =\pi ^2 \left(\frac{1}{6} -\frac{\lvert\nu\rvert}{M} \right) +O\left(\frac{1}{M^2} \right) \label{eq_freffint.3}
\end{equation}
for $\nu\in\mathbb{Z}$ which is independent of $M$, because $\lvert\kappa\rvert$ is sufficiently small. Changing $\nu$ back to $\kappa$ yields the right-hand side of Eq.~\eqref{eq_freffint.1}.

We derive the expression of $S_0$ before that of $S_\nu$. Let us divide $S_0$ into two terms:
\begin{equation}
S_0 =\left(\frac{\pi}{M} \right) ^2 \sum _{\rho =1}^{\frac{M-1}{2}} \left(\frac{1}{\sin ^2 \frac{\pi\rho}{M}} -\frac{1}{\left(\frac{\pi\rho}{M} \right) ^2} \right) +\sum _{\rho =1}^{\frac{M-1}{2}} \frac{1}{\rho ^2} . \label{eq_freffint.4}
\end{equation}
Consider the first term. The fact that the singularity of the function $(\sin x)^{-2} -x^{-2}$ at $x=0$ is removable allows us to replace the sum with the integral:
\begin{align}
\left(\frac{\pi}{M} \right) ^2 \sum _{\rho =1}^{\frac{M-1}{2}} \left(\frac{1}{\sin ^2 \frac{\pi\rho}{M}} -\frac{1}{\left(\frac{\pi\rho}{M} \right) ^2} \right) & =\frac{\pi}{M} \lim _{\epsilon\searrow 0} \int _\epsilon ^{\pi /2} dx\left(\frac{1}{\sin ^2 x} -\frac{1}{x^2} \right) +O\left(\frac{1}{M^2} \right)  \notag \\
& = \frac{\pi}{M} \lim _{\epsilon\searrow 0} \left(\frac{1}{\tan\epsilon} -\frac{1}{\epsilon} +\frac{2}{\pi} \right) +O\left(\frac{1}{M^2} \right) \notag \\
& =\frac{2}{M} +O\left(\frac{1}{M^2} \right) . \label{eq_freffint.5}
\end{align}
Next, the second term on the right-hand side of Eq.~\eqref{eq_freffint.4} is
\begin{align}
\sum _{\rho =1}^{\frac{M-1}{2}} \frac{1}{\rho ^2} & =\sum _{\rho =1}^\infty \frac{1}{\rho^2} -\left(\frac{\pi}{M} \right) ^2 \sum _{\rho =\frac{M+1}{2}}^\infty \frac{1}{\left(\frac{\pi\rho}{M} \right) ^2} \notag \\
& =\frac{\pi ^2}{6} -\frac{\pi}{M} \int _{\pi /2}^\infty \frac{dx}{x^2} +O\left(\frac{1}{M^2} \right) \notag \\
& =\frac{\pi ^2}{6} -\frac{2}{M} +O\left(\frac{1}{M^2} \right) . \label{eq_freffint.9}
\end{align}
We thus have
\begin{equation}
S_0 =\frac{\pi ^2}{6} +O\left(\frac{1}{M^2} \right) . \label{eq_freffint.10}
\end{equation}

Let us now evaluate the sum
\begin{equation}
S_0 -S_\nu =2\left(\frac{\pi}{M} \right) ^2 \sum _{\rho =1}^{\frac{M-1}{2}} \left(\frac{\sin\frac{\pi\nu\rho}{M}}{\sin\frac{\pi\rho}{M}} \right) ^2 . \label{eq_freffint.11}
\end{equation}
We can impose the restriction $\nu\in\mathbb{N}$ without loss of generality since $S_\nu$ is invariant under the transformation $\nu\to -\nu$. We then have
\begin{equation}
S_0 -S_\nu =2\frac{\pi}{M} \int _0^{\pi /2} dx\left(\frac{\sin\nu x}{\sin x} \right) ^2 +O\left(\frac{1}{M^2} \right) =\frac{\pi^2\, \nu}{M} +O\left(\frac{1}{M^2} \right).\label{eq_freffint.12}
\end{equation}
Subtracting this equation from Eq.~\eqref{eq_freffint.10} and using $S_\nu =S_{\lvert\nu\rvert}$ for $\nu\in\mathbb{Z}$ lead to Eq.~\eqref{eq_freffint.3}.

%%%%%%%%%%%%%%%%%%%%%%%%%%
\section{Inverse-Square Decay of the Imaginary-Time Correlation Function} \label{sec_imtcorr}
%%%%%%%%%%%%%%%%%%%%%%%%%%

We evaluate the imaginary-time correlation function $G(\bm{0} ,\rho )$ in Eq.~\eqref{eq_model.19}. Introducing an auxiliary variable $t$ to raise the denominator to the exponent,
we obtain
\begin{equation}
G(\bm{0} ,\rho )\approx\int _0^\infty dt\, e^{-ut} \int _{-\pi}^\pi \frac{d^d \bm{k}}{(2\pi )^d} \exp (-Kt\bm{k}^2)\int _{-\pi}^\pi \frac{d\kappa}{2\pi} \exp (-K_\perp t\kappa ^2 -\pi\alpha t\lvert\kappa\rvert +i\kappa\rho ). \label{eq_imtcorr.2}
\end{equation}
The integral with respect to $\bm{k}$ is written as
\begin{equation}
\int _{-\pi}^\pi \frac{d^d \bm{k}}{(2\pi )^d} \exp (-Kt\bm{k}^2)=\left(\frac{\erf (\pi\sqrt{Kt})}{\sqrt{4\pi Kt}} \right) ^d , \label{eq_imtcorr.3}
\end{equation}
where $\erf z=\frac{2}{\sqrt{\pi}} \int _0^z dt\, e^{-t^2}$ denotes the error function. Because the values at $t\ll u^{-1}$ dominate the integral~\eqref{eq_imtcorr.2}, we keep only the lowest order in $\sqrt{t}$: $\erf (\pi\sqrt{Kt})\approx\sqrt{4\pi Kt}$. Thus,
\begin{equation}
\int _{-\pi}^\pi \frac{d^d \bm{k}}{(2\pi )^d} \exp (-Kt\bm{k}^2)\approx 1. \label{eq_imtcorr.4}
\end{equation}

The integral with respect to $\kappa$ in Eq.~\eqref{eq_imtcorr.2} is reduced to
\begin{align}
\int _{-\pi}^\pi \frac{d\kappa}{2\pi} \exp (-K_\perp t\kappa ^2 -\pi\alpha t\lvert\kappa\rvert +i\kappa\rho ) & =2\Re\int _0^\pi \frac{d\kappa}{2\pi} \exp [-K_\perp t\kappa ^2 -(\pi\alpha t+i\rho )\kappa ] \notag \\
& =\frac{1}{\sqrt{4\pi K_\perp t}} \Re\left[\frac{2}{\sqrt{\pi}} \exp\left(\frac{(\pi\alpha t+i\rho )^2}{4K_\perp t} \right)\int _{(\pi\alpha t+i\rho )/\sqrt{4K_\perp t}}^{(\pi (2K_\perp +\alpha )t+i\rho )/\sqrt{4K_\perp t}} dx\, e^{-x^2} \right] , \label{eq_imtcorr.5}
\end{align}
where we changed variables to $x=\sqrt{K_\perp t} \left(\kappa +\frac{\pi\alpha t+i\rho}{2K_\perp t} \right)$. We can write this integral with the scaled complementary error function $\erfcx z=e^{z^2} (1-\erf z)=\frac{2}{\sqrt{\pi}} e^{z^2} \int _z^\infty dt\, e^{-t^2}$:
\begin{align}
& \quad\,\int _{-\pi}^\pi \frac{d\kappa}{2\pi} \exp (-K_\perp t\kappa ^2 -\pi\alpha t\lvert\kappa\rvert +i\kappa\rho ) \notag \\
& =\frac{1}{\sqrt{4\pi K_\perp t}} \Re\left[\erfcx\frac{\pi\alpha t+i\rho}{\sqrt{4K_\perp t}} -\exp [-\pi ^2 (K_\perp +\alpha )t-i\pi\rho ]\erfcx\frac{\pi (2K_\perp +\alpha )t+i\rho}{\sqrt{4K_\perp t}} \right] . \label{eq_imtcorr.6}
\end{align}
Consider the case of $u\rho\gg\pi (2K_\perp +\alpha )$. Then,
\begin{equation}
\pi\alpha\sqrt{t} \leq\pi (2K_\perp +\alpha )\sqrt{t} \ll\frac{\pi (2K_\perp +\alpha )}{\sqrt{u}} \ll\rho\sqrt{u} \ll\frac{\rho}{\sqrt{t}} \label{eq_imtcorr.7}
\end{equation}
for $t\ll u^{-1}$, and hence we use the series expansion
\begin{equation}
\Re\erfcx\frac{\pi\alpha t+i\rho}{\sqrt{4K_\perp t}} =\Re\erfcx\frac{i\rho}{\sqrt{4K_\perp t}} +\frac{\pi\alpha t}{\sqrt{4K_\perp t}} \Re\erfcx '\frac{i\rho}{\sqrt{4K_\perp t}} +\cdots . \label{eq_imtcorr.8}
\end{equation}
The zeroth-order term
\begin{equation}
\Re\erfcx\frac{i\rho}{\sqrt{4K_\perp t}} =\exp\left(-\frac{\rho ^2}{4K_\perp t} \right)
\end{equation}
is exponentially small for small $t$ and can be neglected, while the first behaves as
\begin{equation}
\Re\erfcx '\left(\frac{i\rho}{\sqrt{4K_\perp t}} \right)\sim\frac{4K_\perp t}{\sqrt{\pi} \,\rho ^2}
\end{equation}
because asymptotically $\erfcx z\sim (\sqrt{\pi} \, z)^{-1}$ as $z\to\infty$ with $\lvert\arg z\rvert <3\pi /4$~\cite{NIST:DLMF:7.12}. We therefore find that
\begin{equation}
\Re\erfcx\frac{\pi\alpha t+i\rho}{\sqrt{4K_\perp t}} \approx\frac{\pi\alpha t}{\sqrt{4K_\perp t}} \Re\erfcx '\frac{i\rho}{\sqrt{4K_\perp t}} \approx\sqrt{4\pi K_\perp t} \,\frac{\alpha t}{\rho ^2} \label{eq_imtcorr.9}
\end{equation}
and likewise
\begin{equation}
\Re\erfcx\frac{\pi (2K_\perp +\alpha )t+i\rho}{\sqrt{4K_\perp t}} \approx\sqrt{4\pi K_\perp t} \,\frac{(2K_\perp +\alpha )t}{\rho ^2} . \label{eq_imtcorr.10}
\end{equation}
Remembering $\rho\in\mathbb{Z}$, we obtain
\begin{equation}
\int _{-\pi}^\pi \frac{d\kappa}{2\pi} \exp (-K_\perp t\kappa ^2 -\pi\alpha t\lvert\kappa\rvert +i\kappa\rho )\approx\frac{\alpha t-(-1)^\rho (2K_\perp +\alpha )t\exp [-\pi ^2 (K_\perp +\alpha )t]}{\rho ^2} . \label{eq_imtcorr.11}
\end{equation}

Substituting Eqs.~\eqref{eq_imtcorr.4} and \eqref{eq_imtcorr.11} into Eq.~\eqref{eq_imtcorr.2} and calculating the integral result in
\begin{equation}
G(\bm{0} ,\rho )\approx\frac{\alpha}{(u\rho )^2} -(-1)^\rho \frac{2K_\perp +\alpha}{[\{ u+\pi ^2 (K_\perp +\alpha )\}\rho ]^2} . \label{eq_imtcorr.12}
\end{equation}
Since the first term is the leading contribution near the criticality, we finally arrive at Eq.~\eqref{eq_model.23}.

\bibliography{paper1}

\end{document}